\documentclass[aps,nofootinbib]{revtex4-2}
\usepackage{graphicx}
\usepackage{amssymb}
\usepackage{graphicx}
\usepackage{amsmath}
\usepackage{eurosym}
\usepackage{color,graphicx}
\usepackage{dsfont}
\usepackage{amsmath}
\usepackage{amssymb}
\usepackage{amsthm}
\usepackage{stmaryrd}
\usepackage{mathrsfs}
\usepackage{textcomp}
\usepackage{amsfonts}
\usepackage{txfonts}
\usepackage{subfig}

\begin{document}

\title{Refining Bounds for Snyder and GUP Models through Seismic Wave Analysis}

\author{Aleksander Kozak}
\email{aleksander.kozak2@uwr.edu.pl}
\affiliation{Institute of Theoretical physics, University of Wroclaw, pl. Maxa Borna 9, 50-206 Wroclaw, Poland\\ and \\
Universidad Cat\'olica del Norte, Antofagasta, Chile}

\author{Anna Pacho\l}
\thanks{Corresponding author}
\email{anna.pachol@usn.no}
\affiliation{Department of Microsystems, University of South-Eastern Norway, Campus Vestfold, Norway}

\author{Aneta Wojnar}
\email[E-mail: ]{aneta.wojnar2@uwr.edu.pl}
\affiliation{Departamento de F\'isica Te\'orica \& IPARCOS, Universidad Complutense de Madrid, E-28040, 
Madrid, Spain \\
and\\
Institute of Theoretical physics, University of Wroclaw, pl. Maxa Borna 9, 50-206 Wroclaw, Poland}

\begin{abstract}
This study investigates possibility of placing bounds on the parameters, arising from the non-commutative Snyder space-time model and Generalized Uncertainty Principle (GUP) approach, by utilizing seismic data. We investigate the dependence of constraints  on the type of realization used for the quantum phase space. Results indicate improved bounds compared to prior studies, with the model parameter $\beta_0$ constrained to be less than $5.2\times 10^{44}$ for certain choice of realizations. This approach demonstrates the potential for using Earth's empirical data to refine constraints on GUP parameters.
\end{abstract}

\maketitle

\section{Introduction}
The Earth, Moon, and other astrophysical bodies provide natural laboratories for exploring fundamental physics \cite{luo2012detectable, garcia2019lunar,fedderke2021earth,leane2021exoplanets,raffelt1996stars}, thanks to the knowledge we have of their internal structures. For example, the capture of hypothetical dark matter particles by these objects could result in particle annihilation and heat release, influencing their thermal behavior. Measuring this additional heat provides a means to constrain the properties of dark matter. Moreover, alternative theories to General Relativity (GR) predict differences in the cooling rates of astrophysical bodies, offering possible observational tests for these models \cite{Benito:2021ywe, Wojnar:2021xbr,Kalita:2022trq}. The Moon’s lack of internal heat and the predictable cooling patterns of white dwarfs present valuable opportunities for probing exotic particles and testing extensions or modifications to Einstein's gravity \cite{garani2020constraints,isern2022white}. These naturally occurring conditions complement laboratory experiments by offering insights into particle interactions and possible modifications of GR.

At the same time, seismology—particularly asteroseismology—has emerged as a powerful tool for investigating fundamental physics. Asteroseismology involves analyzing brightness fluctuations in stars to study their internal structures and dynamics \cite{aerts2010asteroseismology}. These variations reveal resonant oscillations, such as pressure modes and gravity modes, which provide valuable information about stellar evolution and structure. Acoustic modes (or pressure), for instance, typically arise in the convective envelopes of low-mass stars (such as solar-like stars and red giants) \cite{de2009non}, while gravity modes occur in the radiative envelopes of massive stars. These oscillations enhance our understanding of nuclear processes, chemical composition, convection, rotation, and magnetic fields within stars \cite{andersson1996gravitational,benhar2004gravitational}.

Furthermore, stellar and planetary oscillations have proven to be a gateway for exploring phenomena like dark matter, dark energy, and modifications to GR. The oscillation frequency spectra of stars and neutron stars, for example, can be used to constrain their mass and radius, leading to insights into the nuclear equation of state. Oscillations may also carry imprints of scalar fields associated with dark energy or inflation and could reveal exotic particles \cite{casanellas2013first,mendes2018new}. Asteroseismic models of solar-like stars have been crucial in refining exclusion limits for dark matter and testing deviations from Einstein’s theory of gravity. Additionally, seismic observations have been utilized to investigate time variations in fundamental constants \cite{bellinger2019asteroseismic}, including the gravitational constant. By comparing observed oscillation frequencies with theoretical predictions, researchers can constrain changes in these constants over cosmic timescales.

In the specific case of the Sun, its detailed spectrum of oscillations offers a precise probe into its interior, providing unique opportunities to detect fifth force or deviations from GR \cite{saltas2019obtaining,saltas2022searching}. Overall, the study of seismic phenomena in stars and planets serves as a natural laboratory for advancing our understanding of fundamental physics, bridging the gap between observational data and theoretical models. In this paper, we are interested in the use of seismic data from Earth in an application to probe models arising in quantum gravity phenomenology and in the so called Generalized (Heisenberg) Uncertainty Principle (GUP) framework.

Quantum gravity phenomenology explores the effects of quantum gravity at energy scales that are currently beyond the reach of direct experimental methods. 
At extremely high energy scales, where quantum effects become as important as gravity, the conventional framework of general relativity breaks down and continuous space-time description is no longer applicable. This regime is characterized by the Planck length $L_P\sim \sqrt{\frac{\hbar G}{c^3}}$. Quantum gravity theories (such as string theory, loop quantum gravity, modified gravity and others) attempt to describe the behavior of space-time, and gravitational and quantum interactions under these extreme conditions. Direct experimental tests  at the Planck scale are currently impossible due to technological limitations. The energy required to probe such scales is far beyond the capabilities of modern particle accelerators. Nevertheless, through quantum gravity phenomenology, instead of directly testing quantum gravity effects, we construct simplified and effective models that capture some of the anticipated effects at lower energy scales. 
Such phenomenological models often involve modifying the classical solutions, to account for the potential new quantum gravity effects. These modifications arising from various quantum gravity approaches can be studied in the effective way and then be compared against existing data or measurable physical effects. 
One of such phenomenological approaches is the GUP \cite{Kempf:1994su,kempf1995hilbert,Maggiore:1993rv,Maggiore:1993zu,Chang:2001bm,Chang:2001kn,Bosso:2023aht} where one assumes that the Heisenberg’s uncertainty principle becomes modified, giving an effective way to investigate possible new effects arising from such modifications in cosmology, astrophysics, statistical mechanics just to mention few applications. One treats these as phenomenological models, and the results of GUP approach often offer constraints on the model parameter ($\beta_0$) which governs the introduced modification. The recent review with an overview of possible bounds has been presented, for example in \cite{Bosso:2023aht}, where the upper bounds listed range between $10^{16}$ - $10^{90}$ and were obtained in various considerations (in  gravitational experiments and observations as well as in the tabletop experiments not
related to gravity).

In this paper, we base on the non-commutative (NC) geometry describing the quantum structure of space-time resulting in the modification of the quantum phase space together with GUP framework, to study the implications of these modifications on the possible physical effects. Our study gives more stringent constraints on the model parameter compared to other works that analyze astrophysical objects, but here we consider Earth as such object. In fact considering seismological data from Earth gives bounds much closer to the upper bounds obtained in the tabletop experiments not related to gravity (see \cite{Bosso:2023aht} and references therein).
To this aim we use a novel approach \cite{kozak2021metric,Kozak:2023axy,kozak2023earthquakes} which relies on the use of seismic data from Earth to constrain an additional term appearing in the effective Poisson equation resulting from the modifications arising in quantum phase space due to the noncommutativity of space-time \cite{pachol2023constraining}. These corrections, being second-order (therefore they will appear in the Poisson equation), are relatively small but have been shown to influence astrophysical structure equations and microphysics. For rocky planets, their interior layers are well-described by a modified polytropic model \cite{Seager:2007ix}, which incorporates quantum effects \cite{salpeter1967theoretical}. In contrast, the outer layers are effectively characterized by the empirical Birch law.

The quantum space-time we use is given by the Snyder model \cite{snyder1947quantized}, where the space-time coordinates are NC and the commutation relation between them is proportional to the Lorentz generators. The Lorentz symmetry underlying this quantum space-time remains undeformed at the algebraic level hence the Snyder model is an example of a Lorentz-covariant NC space-time, featuring a fundamental length scale.
The deformation of the quantum-mechanical phase space we consider, up to the linear order in the non-commutativity parameter $\beta$, has the following form \cite{meljanac2021heisenberg} \footnote{reduced to the relation between spacial coordinates, in accordance with the GUP framework.}:
\begin{eqnarray}
[p_{i},\hat{x}_{k}] &=&-i\hbar\delta_{ik}\left( 1+\beta \left( \chi-\frac{1}{2}\right) p^{j}p_j\right) -2i\hbar\chi\beta p_{i}p_{k}+O(\beta ^{2}), \label{gen_real_p-x_sp}
\end{eqnarray}
where we rely on the most general realization for the non-commuting Snyder coordinates \cite{battisti2010scalar,meljanac2021heisenberg,Pachol:2023tqa} which are parametrized by $\chi$.

By using such general form of the modified phase space, we can analyze the phenomenological predictions, specifically the relation between the effects related to the extra terms introduced in the phase space in this model\footnote{The modified phase space measure based on the Liouville theorem
is $\frac{d^3xd^3p}{1+\Omega p^2}$, where $\Omega=\beta(4\chi-\frac{3}{2})$. To obtain this we have relied on the results of \cite{Chang:2001bm} and we refer the reader to the Appendix in our previous work \cite{pachol2023constraining} for the full details and approximations used. We point out that the value of $\Omega$ is related with the choice of the realization
parameter $\chi$ appearing in the Snyder phase space we
consider as well as the noncommutativity parameter $\beta$.} and the matter description arising from analyzing the seismological data from Earth. Additionally, by considering the one-parameter family of modified phase spaces of the Snyder model, we can investigate if any measurable effects favor one realization (i.e. a specific value of the parameter $\chi$) over others and provide new bounds on the non-commutativity parameter $\beta$.

\section{
Effective Poisson equation and Earth model}

One can show \cite{pachol2023constraining} that the deformed phase space (\ref{gen_real_p-x_sp}) leads to the modified Poisson equation for the gravitational potential $\phi$
\begin{equation}\label{poisson2}
   \nabla^2\phi =  4\pi G \rho - \tilde\epsilon \nabla^2 \rho^\frac{4}{3},
\end{equation}
where $G$ is the Newton constant and $\rho$ an energy density of an astrophysical body. In what follows, we will consider the spherical-symmetric case such that $\phi=\phi(r)$ and $\rho=\rho(r)$ depend solely on the radial coordinate. 
The additional term appears due to the fact that the modification in the
phase space volume leads to the corrections arising in the partition function, which in turn modify the Fermi-Dirac equation of state (for details, see Sec. 3 in  \cite{Pachol:2023tqa}):
\begin{equation}\label{pres5}
    P_{T\to 0} =K_1{\rho}^\frac{5}{3}\left[
    1-\varepsilon{\rho}^\frac{2}{3}
    \right],
\end{equation}
where $K_1= \frac{2}{5}vu^\frac{5}{2}\mu_e^{-\frac{5}{3}}$ and we have defined $v=\frac{(2m_e)^\frac{2}{3}}{3\pi^2 \hbar^3}$ and $u=(3 \pi^{2} \hbar^{3} N_A)^\frac{2}{3}/2m_e$, while the remaining constants have the standard meaning. On the other hand, the parameter $\varepsilon=\frac{3}{7}(\frac{3\pi^2\hbar^3N_A}{\mu_e})^{\frac{2}{3}}\Omega=4.47878\times10^{-52}\Omega$, where $\Omega$ depends on different realizations of the Snyder model (cf. footnote 2). 
The parameter $\tilde\epsilon$ appearing in the modified Poisson equation is defined as $\tilde\epsilon=\frac{49}{20} K_1 \varepsilon \rho_c^{\frac{2}{3}} $, where $\rho_c$ is the core density of an astrophysical object. Such parameterized effective Poisson equation \eqref{poisson2} has been used to describe internal processes happening in low-mass stars, allowing to constrain the model to the orders of magnitude much better than with the use of the compact objects and astrophysical events \cite{Bosso:2023aht}. 
Moreover, one can show that the noncommutativity correction appearing in \eqref{poisson2} within the non-relativistic limit can be also interpreted as a "modified gravity" correction to the Poisson equation \eqref{poisson2}, with a not modified polytropic equation of state, for more details see \cite{pachol2023constraining}.

In contrast to compact and stellar objects, where uncertainties arise from equations of state and atmospheric properties \cite{Kozak:2022hdy,Gomes:2022sft,wojnar2023fermi}, Earth seismology offers valuable insights into the planet's interior \cite{poirier2000introduction,dziewonski1981preliminary,kennett1991traveltimes}. Combining seismic data with precise measurements of Earth's mass and moment of inertia, we gain a powerful tool to constrain gravity and GUP models. This approach leverages well-understood physics and helps mitigate uncertainties associated with model assumptions. Moreover, recent advancements in seismographic tools \cite{butler2021antipodal,frost2019orientation,pham2023up} and laboratory experiments simulating extreme temperatures and pressures in Earth's interior have significantly enhanced our understanding of Earth's interior properties, particularly those of iron and its compounds \cite{merkel2021femtosecond}. Additionally, new neutrino telescopes provide information on density, composition, and abundances of light elements in the outer core, further reducing uncertainties related to Earth's core characteristics \cite{winter2007neutrino,donini2019neutrino}.

However, concerns have been raised about the seemingly negligible impact of  gravity effects in stellar and planetary physics. Although the gravitational effects may only result in small changes to layer densities and thicknesses, they remain significant \cite{Kozak:2021zva,Kozak:2021fjy,Wojnar:2021xbr}. Fortunately, our extensive and accurate knowledge of the Solar System planets, especially Earth \cite{bills1999obliquity,williams1994contributions} enables us to utilize available data to constrain the theories introducing modifications, as in for example the modified Poisson equation above \cite{kozak2021metric,Kozak:2023axy,kozak2023earthquakes}. Through a simplified approach, we have shown \cite{Kozak:2023axy,kozak2023earthquakes} that one can achieve accuracy up to the $2\sigma$ level in constraining the theory parameters in the context of modified gravity approach. Here, we will focus on the NC Snyder model and GUP approach instead.

Apart from the effective Poisson equation \eqref{poisson2}, in order to describe a terrestrial planet in the hydrostatic equilibrium, we need the hydrostatic equilibrium equation given by
\begin{equation}\label{hydro}
    \frac{d\phi}{dr}=-\rho^{-1}\frac{dP}{dr},
\end{equation}
where $P$ is pressure, and in our approach we will consider it as barotropic, that is, it is a function of density only, $P=P(\rho)$. Earth's mass $M$, enclosed within a ball with radius $r=R$, takes the standard form in the non-relativistic limit
\begin{equation}\label{mass}
     M=\int_0^R 4\pi\tilde r^2 \rho(\tilde r) d\tilde r.
\end{equation}
To simplify the analysis, we assume adiabatic compression, meaning that there is no heat exchange between Earth's layers. The planet is also assumed to have radially symmetric shells featuring a given density jump $\Delta\rho=600$ between the inner and outer core, a central density $\rho_c=13050$, and a density at the mantle's base $\rho_m=5563$ (all values in kg/m$^3$).

Therefore, since for pressures $P> 10^4$ GPa one needs to take into account the electron degeneracy, we will describe the Earth's innermost layers with the equation of state for non-relativistic fermions, that is, given by a simple polytrope as $P=K_1\rho^{5/3}$.
On the other hand, the densities in the outer layers follow the empirical Birch's law
\begin{equation}
    \rho = a + b \mathrm{v}_p,
\end{equation}
where $a$ and $b$ are parameters that depend on the mean atomic mass of the material in the upper mantle \cite{dziewonski1981preliminary}. The longitudinal elastic wave $\mathrm{v}_p$ and the transverse elastic wave $\mathrm{v}_s$ allow us to define the seismic parameter $\Phi_s$ \cite{poirier2000introduction}:
\begin{equation}\label{seismic}
 \Phi_s= \mathrm{v}_p^2 - \frac{4}{3} \mathrm{v}_s^2.
\end{equation}
These velocity-depth profiles, as shown in Fig. 1 in \cite{Kozak:2023axy}, are derived from travel-time distance curves for seismic waves and periods of free oscillations \cite{poirier2000introduction,bolt1982inside, bullen1985introduction}. They provide pressure, density, and elastic moduli profiles as functions of depth. The seismic parameter \eqref{seismic} is related to the bulk modulus $K$ (incompressibility) as
\begin{equation}
   \Phi_s= \frac{K}{\rho}. 
\end{equation}
By applying the definition of the bulk modulus $ K= \frac{dP}{d \mathrm{ln}\rho}$, we can write $\Phi_s= \frac{dP}{d\rho}$.  Thus, the seismic parameter includes information on the equation of state, which we can use in Eq. \eqref{hydro} to express
\begin{equation}
    \frac{d\rho}{dr}=-\rho \Phi_s^{-1} \frac{d\phi}{dr}. 
\end{equation}
The mass equation \eqref{mass} and the polar moment of inertia 
\begin{equation}
   I= \frac{8}{3}\pi \int_0^R r^4\rho(r) dr 
\end{equation}
serve as constraints with high-accuracy observational values \cite{luzum2011iau,chen2015consistent}.

\section{Numerical approach and results}

\begin{figure}[t]
\centering
\includegraphics[scale=.75]{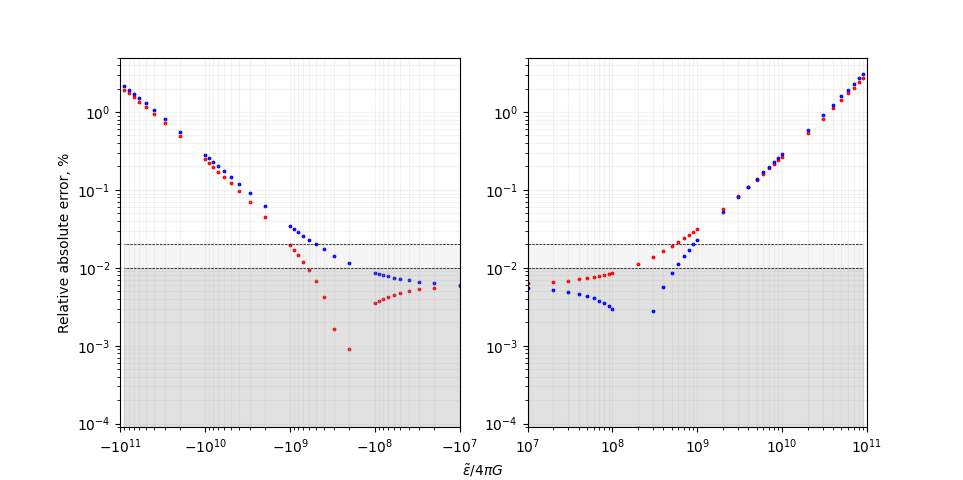}
\caption{[color online] The relative absolute errors for Earth's mass and moment of inertia. Red dots represent moment of inertia errors, while blue dots represent mass errors. The dark grey stripe represents the 1-sigma region, and the light grey denotes the 2-sigma region. The grey region encompasses uncertainties for both mass and moment of inertia, as their ratios of sigma to mean value are similar (approximately 0.01\%). The unit of $\tilde{\epsilon}$ is $\text{m}^6\: \text{kg}^{-4/3}\: \text{s}^{-2}$. }
\label{results}
\end{figure}

To further advance our study, we rely on data provided by \cite{dziewonski1981preliminary} and related references, which contain measured seismic wave velocities. We calculate density profiles by assuming values for the free parameters of the PREM model, namely the mantle density $\rho_m$, core density $\rho_c$, and density jump between the inner and outer core $\Delta\rho$. Our aim is to obtain density values consistent with those predicted by PREM while accounting for uncertainties arising from measurements. To achieve this, we utilize a Python script to perform the calculations. We vary the values of $\tilde\epsilon$ ({appearing in the modified Poisson equation \eqref{poisson2} and related with the non-commutativity and Snyder realizations parameters}) and observe their effects on the calculations. For a more straightforward analysis, we employ a simplified model to assess crucial parameters and determine the order of magnitude of $\tilde\epsilon$ where effects from introduced modifications (in quantum phase space and GUP) align with observed constraints. The results are presented in Fig. \ref{results}. {The parameter $\tilde\epsilon$ is rescaled for computational purposes by $4\pi G$ and its different values are represented on the horizontal axis. The deviation of the resulting mass and polar moment of inertia of the Earth from the measured values is placed on the vertical axis; it is simply the absolute relative error. The shaded areas correspond to $1\sigma$ and $2\sigma$ uncertainty regions about the experimental values, which, expressed as percentage errors, are similar for both the mass and the moment of inertia (they are equal to approximately  $0.01\%$ in the case of $1\sigma$ deviations). }

\begin{figure*}
  \centering
  \advance\leftskip-1.5cm
  \subfloat{\includegraphics[scale=0.5]{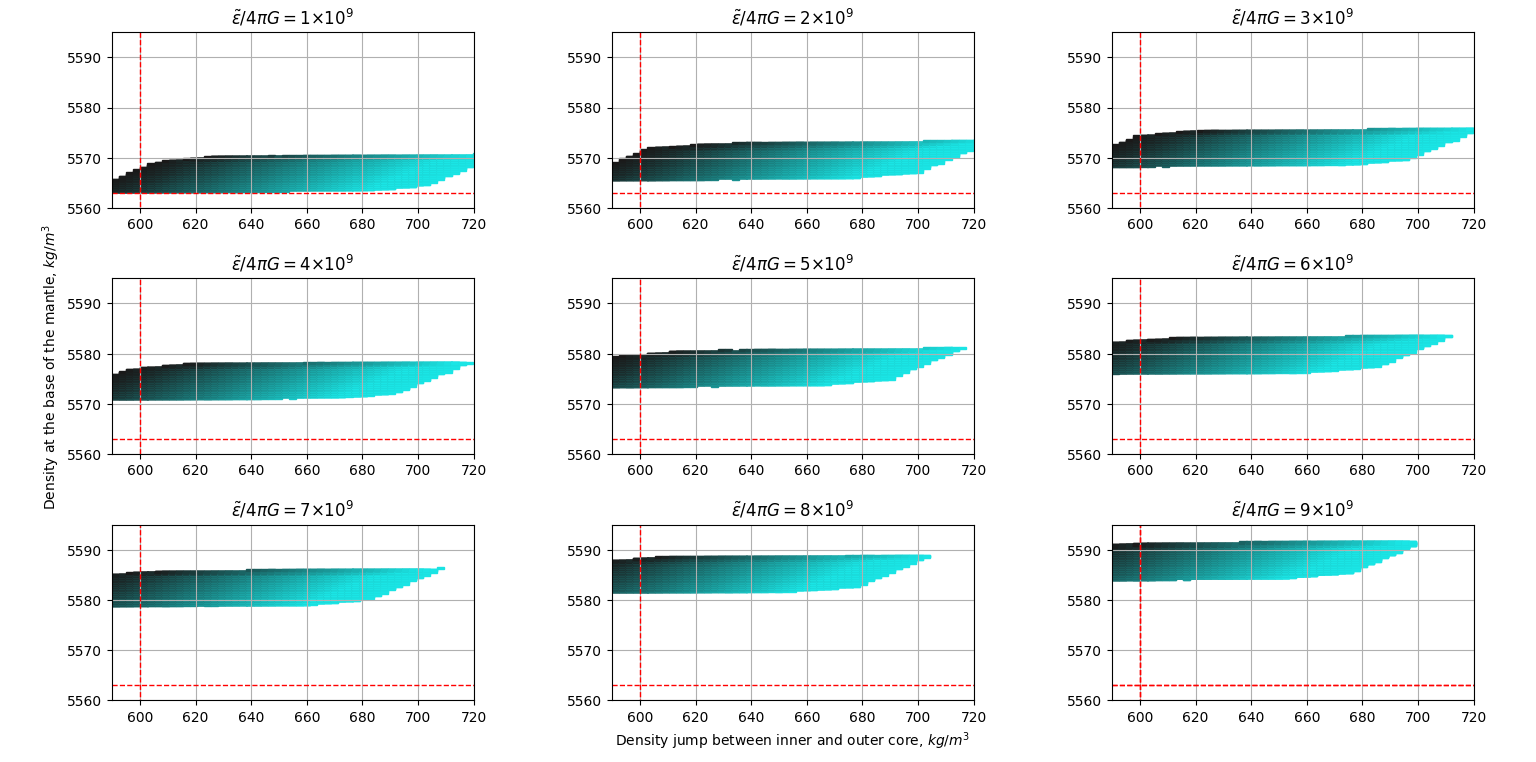}} \\
  \subfloat{\includegraphics[scale=0.7]{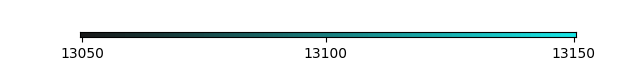}}
  \caption{[color online]  $2\sigma$ confidence regions of the theory parameters $(\rho_c, \rho_m, \Delta \rho)$ for different values of the $\epsilon=\tilde\epsilon/4\pi G$ parameter, being of order $10^9  \text{m}^3 \: \text{kg}^{-1/3}$. The darker color corresponds to lower values of the central density, while the brighter one - to higher. The range of the central density is shown in the color bar below the figures. The units are kg/m$^3$. The red dashed lines correspond to the PREM values of the density jump and the density at the base of the mantle.} 
  \label{1e9}
\end{figure*}

\begin{figure*}
  \centering
  \advance\leftskip-1.5cm
  \subfloat{\includegraphics[scale=0.5]{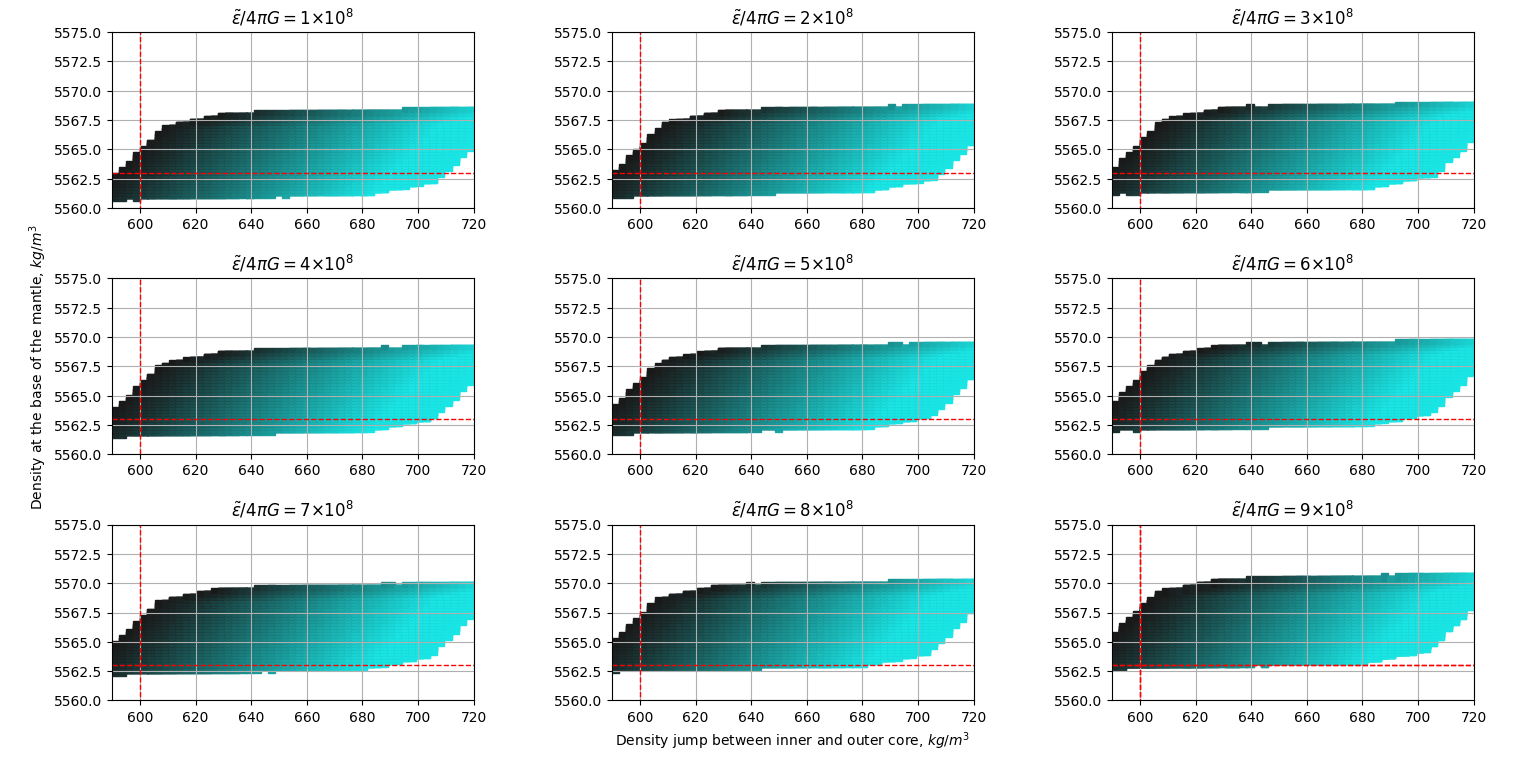}} \\
  \subfloat{\includegraphics[scale=0.7]{pasek.png}}
  \caption{[color online]  $2\sigma$ confidence regions of the theory parameters $(\rho_c, \rho_m, \Delta \rho)$ for different values of the $\epsilon=\tilde\epsilon/4\pi G$ parameter, being of order $10^{10} \text{m}^3 \: \text{kg}^{-1/3}$. The darker color corresponds to lower values of the central density, while the brighter one - to higher. The range of the central density is shown in the color bar below the figures. The units are kg/m$^3$. The red dashed lines correspond to the PREM values of the density jump and the density at the base of the mantle.} 
  \label{1e8}
\end{figure*}

It is important to note that unlike the PREM model, the value of the central density is not a result of solving differential equations but rather an initial assumption. Additionally, we focus on modeling the interior layers since we expect corrections introduced to have weaker effects in the outermost regions. To this end, we assume that Birch's law holds, and all density values are directly taken from the PREM model. We then integrate \eqref{poisson2} together with the mass relation \eqref{mass} for different values of the parameters of the model. The goodness of fit for a particular parameter choice is determined by how well the calculated Earth's mass and polar moment of inertia agree with the measured values: $M_\oplus = (5.9722 \pm 0.0006)\times 10^{24} \text{kg}$ \cite{luzum2011iau} and $I_\oplus = (8.01736\pm 0.00097)\times 10^{37} \text{kg m}^2$ \cite{chen2015consistent}.

Our study reveals that the following bounds can be placed on the considered parameters, ensuring that the deviations of Earth's mass and polar moment of inertia do not exceed $2\sigma$:
$$-0.4191 \leq \tilde{\epsilon} \leq 0.5023 \quad\mbox{m}^6\: \mbox{kg}^{-4/3}\: \mbox{s}^{-2}$$ or, alternatively, 
\begin{equation}\label{alphaomega}
- 3.3823\times 10^{42}\leq \Omega \leq 4.0588\times 10^{42}\quad (\mbox{kg m s}^{-1})^{-2}    
\end{equation}
obtained for the Earth's central density $\rho_c=13050$ kg/m$^3$. Note 
$\Omega=\beta(4\chi-\frac{3}{2})$ is related with the modification arising in the measure of the phase space, see footnote 2.

When considering PREM as a valid Earth model, the associated uncertainties in the moment of inertia and mass establish constraints on its parameters. Nonetheless, due to PREM's inherent imperfections, deviations in density parameters may arise from our initial assumptions. Notably, there exists a parameter space within a given theoretical framework where all three density parameters align with experimental measurements, as elaborated in our analysis \cite{Kozak:2023axy}. Among these parameters, $\rho_m$ exhibits a more limited range of variation compared to $\Delta\rho$ and $\rho_c$.

For instance, when $\epsilon = \frac{\tilde{\epsilon}}{4\pi G}=10^8 \text{ m}^3\:\text{kg}^{-1/3}$, the alterations required in $\rho_m$ to maintain consistent mass and polar moment of inertia, as compared to the scenario with $\epsilon=0$, are exceedingly small, measuring just $0.04\%$. In contrast, the most extreme uncertainty in the PREM model, involving deviations of approximately $50  \text{ kg m}^{-3}$ while holding $\Delta\rho$ and $\rho_c$ constant, amounts to $0.9\%$. Under such deviations, the parameter $\epsilon$ experiences an increase of nearly two orders of magnitude, as depicted in Figures \ref{1e9} and \ref{1e8}. This illustrates that the influence of altering the theoretical parameter $\epsilon$ on $\rho_m$ is dwarfed by the uncertainties inherent to the PREM model, suggesting that the choice of parameter effect on the outcomes remain relatively minor in comparison to the uncertainties associated with the Earth model itself.

Consequently, since the parameters of {non-commutativity $\beta$ and Snyder realization $\chi$} (appearing in the $\Omega$ terms) exert a notable impact on the permissible range of $\rho_m$, enhancing the precision of $\rho_m$ by adopting a more refined Earth model will further enhance our capacity to constrain them.
It is worth noting that in this study, we constrain the {Snyder and GUP} models by using the density parameters derived from PREM, operating under the assumption that PREM serves as a plausible representation of the Earth \cite{Kozak:2023axy}.

Now let us discuss the implications on the bounds for the non-commutativity parameter $\beta$. For the realizations of the Snyder model for which $(4\chi-\frac{3}{2})>0$ (i.e. when $\chi>0.375$) we obtain, from \eqref{alphaomega}, the following:
\begin{equation}\nonumber
\beta\leq \frac{4.0588\times 10^{42}}{(4\chi-1.5)} \text{\, (kg m s}^{-1})^{-2}.
\end{equation}
For example choosing the value of $\chi=0.5$ of the (original \cite{snyder1947quantized}) Snyder realization, we get the following bound:
$$\beta\leq 8.12\times 10^{42}\text{\, (kg m s}^{-1})^{-2} .$$ 
Considering the dimensionless parameter $\beta_0=\beta M^2_Pc^2=2.3\times 10^2\beta$ our bound then is:
\begin{equation}\label{beta_0chiplus}
\beta_0 \leq 1.9\times 10^{45}.
\end{equation}
Choosing values of $\chi\approx 1$ we can reduce the order of magnitude by 1 already, i.e. obtain $ \beta_0 \leq  10^{44}$. Choosing values of $\chi\approx 3$ we get further reduction of the order $ \beta_0 \leq  10^{43}$. 

On the other hand, for $(4\chi-\frac{3}{2})<0$ (i.e. $\chi<0.375$) we get:
\begin{equation}\nonumber
\beta \leq \frac{3.3823\times 10^{42}}{(1.5-4\chi) } \text{\, (kg m s}^{-1})^{-2}.
\end{equation}
In this case one of the distinguished values of the parameter $\chi$ is $\chi=0$ (which was considered in e.g. \cite{Maggiore:1993rv},\cite{Maggiore:1993zu} by Maggiore and studied in many GUP related effects). For this value we get: 
$$\beta\leq 2.25\times 10^{42} \text{\, (kg m s}^{-1})^{-2},$$
\begin{equation}\label{beta_0chiminus}
\beta_0\leq 5.2\times 10^{44}.
\end{equation}
We see that for example for the 'Maggiore' realization ($\chi=0$) the bound we obtain (\ref{beta_0chiminus}) is slightly better than, for the original Snyder realization ($\chi=\frac{1}{2}$) .

Nevertheless, it seems that our investigation provides upper bounds of order $10^{44}$ which are better than in our previous work in the case of low-mass stars ($10^{48}$) \cite{pachol2023constraining} and in broader literature so far, up to our knowledge, see e.g. \cite{Bosso:2023aht} where comprehensive classification of the existing constraints up to date is provided (Table 2 therein). Considering seismological data from Earth gives bounds much closer to the upper bounds obtained in the tabletop experiments not related to gravity, see Table 1 therein. 

\section{Conclusions}

 The primary objective of this study was twofold: first, to explore the applicability of the recently introduced method proposed by \cite{kozak2021metric,Kozak:2023axy,kozak2023earthquakes} for testing modifications arising from non-commutativity of space-time; and second, to assess whether this method yields improved constraints compared to existing ones \cite{Bosso:2023aht}. Building upon our previous research \cite{pachol2023constraining}, we focused our investigation on the non-commutative space-time and GUP approach, using the Snyder model as an illustrative example. Our findings demonstrated that, in line with our expectations, this approach yielded analogous physical outcomes across various choices for the realizations of the deformed phase space. Although here, as opposed to our previous results from \cite{pachol2023constraining} and \cite{Pachol:2023tqa}, both ranges for the parameter $\chi$, i.e. $\chi>0.375$ (positive $\Omega$) and $\chi<0.375$ (negative $\Omega$) produce bounds. Out of the two, slightly better bounds are obtained for $\chi<0.375$ (including the case of $\chi=0$ for the 'Maggiore' realization of the Snyder model). One can note that further decreasing the values of $\chi$ leads to the reduction of orders of magnitude for the bound of the non-commutativity parameter $\beta_0$.

 Note that the quantum gravity corrections would potentially affect microphysics in our scenario - the equation of state and other thermodynamical properties of the astrophysical objects. This type of corrections are quite commonly investigated in the GUP approach in astrophysical context, here we have focused on data from Earth instead.
 However, it is worth to mention that the same corrections can be interpreted as terms resulting from \textit{some} modified gravity model, and on level of the classical equations describing a star or a planet, those effects are indistinguishable from each other. Because of that fact we can use various methods developed by the modified gravity community to test and to constrain the quantum space-times proposals.

In our analysis, we employed seismic data \cite{dziewonski1981preliminary}, which comprises information on the velocities of longitudinal and transverse elastic waves as well as the depths of their propagation. These elastic waves carry valuable insights into the material properties within the Earth, which can be decoded through the seismic parameter incorporated into the (modified) hydrostatic equilibrium and Poisson equations. By solving these equations under various theory parameters and considering different boundary and initial conditions, such as central densities and density jumps between layers, we derived density profiles that deviate from the preliminary reference Earth model \cite{dziewonski1981preliminary}, originally constructed based on Newtonian gravity.

In contrast, Earth's observable attributes, such as its mass and moment of inertia, impose stringent constraints on these density profiles and, consequently, on $\Omega$ combination of parameters governing non-commutativity and the choice of phase space realization. Despite the simplicity of our model, it proved capable of constraining the Snyder non-commutativity parameter $\beta$ using the Earth's empirical data. As mentioned, we obtained improved bounds with respect to previous works, that is, 
$ \beta_0 \leq 5.2\times 10^{44}$ for the case when $\chi < 0.375$ and $\beta_0\leq 1.9\times 10^{45}$ for $\chi > 0.375$, and both possible ranges for the realizations lead to obtaining constraints (which was not the case in our previous work).

While our studies have yielded more stringent constraints on the non-commutative parameter compared to other works that analyze astrophysical objects, it's important to note that our approach has inherent limitations due to underlying assumptions and simplifications. Foremost among these is the assumption of spherical symmetry, a point previously discussed in the text. Earth's shape deviates from a perfect sphere, and the moment of inertia is influenced by rotation and the particular symmetries it induces. This issue is evident in PREM, which fails to produce a moment of inertia consistent with observational values within its precision. To address this challenge without introducing the complexities of Earth's true geometry, one potential solution is to estimate the equatorial moment of inertia relative to the polar moment by applying travel time ellipticity corrections to PREM. Expressions for ellipsoidal correction of travel time can be found in \cite{kennett1996ellipticity,kennett1991traveltimes}.

Furthermore, both PREM and our models are one-dimensional and assume spherical layers. Recognizing their imperfections and accounting for variable density jumps will undoubtedly introduce non-negligible effects on the moment of inertia and mass. Additionally, the assumption of adiabatic compression, which entails a constant temperature with depth, demands consideration in our future work.

Moreover, PREM fails to account for travel times of seismic waves that probe the boundaries of the outer and inner core, rendering it unsuitable for body wave studies in these regions, as adopted in our simplified approach. Instead, a more precise model like AK135-F \cite{kennett1995constraints,montagner1996reconcile}, which incorporates the intricacies of core waves, should be employed. Alternatively, equations of state for modeling core density and bulk moduli \cite{irving2018seismically} could replace reliance on seismic data, which may be subject to uncertainties in density jumps at the inner and outer core boundaries.

Concerns may arise regarding the use of Birch's law for the outer layers within {considered here non-commutative model}. As previously mentioned, this is an empirical law with coefficients obtained experimentally. Although gravity offers insights into material properties, such as chemical potential \cite{kulikov1995low}, chemical reaction rates \cite{lecca2021effects}, specific heat \cite{farag2014towards,riasat2023effect,verma2019effect}, Debye temperature, crystallization processes \cite{Kalita:2022trq}, and equations of state \cite{kim2014physics,wojnar2023fermi,Pachol:2023tqa}, its application in this context remains justified. However, it's worth noting that the coefficients of Birch's law may need reevaluation when dealing with seismic data from Mars, considering that the composition of outer layers varies among terrestrial planets.

In conclusion, findings in this paper highlight the significant potential of utilizing seismological data from Earth (considered with the above mentioned simplifications) to test fundamental physics, particularly in the realms of modified and quantum gravity. Such simplified models allow us to focus on the core physics of the phenomena under investigation by reducing the complexity associated with higher-dimensional systems. By adopting a novel approach that leverages seismic data to constrain modifications in the effective Poisson equation arising from noncommutative space-time, bounds on theoretical parameters have been achieved that compete—and, in some cases, surpass—those obtained through traditional tabletop experiments. Remarkably, the seismic method introduced in \cite{Kozak:2023axy}, and used here, has demonstrated an efficacy 40 orders of magnitude greater than the most recent cosmological data in constraining specific gravity models \cite{Gomes:2023xzk}.\newline
A crucial improvement for this approach would lie in incorporating Earth's rotation, which would refine the model-predicted polar moment of inertia. Further advancements will require a deeper understanding of seismic waves propagating through Earth's core and the use of more precise equations of state, albeit with some associated uncertainties. This methodology emphasizes the growing importance of seismic data as a transformative tool for probing the intricate connection between Earth's internal dynamics and the fundamental laws governing the universe.

\section*{Acknowledgements}
AK acknowledges financial support from Fondecyt de Postdoctorado 2025, no. 3250036.
AP has been supported by the Polish National Science Center (NCN), project UMO-2022/45/B/ST2/01067.
 AW acknowledges financial support from MICINN (Spain) {\it Ayuda Juan de la Cierva - incorporac\'ion} 2020 No. IJC2020-044751-I and from the Spanish Agencia Estatal de Investigaci\'on Grant No. PID2022-138607NB-I00, funded by MCIN/AEI/10.13039/501100011033, FEDER, UE, and
ERDF A way of making Europe. 

 The authors acknowledge COST Actions CaLISTA CA21109, BridgeQG CA23130 and FuSe CA24101.

\end{document}